\documentclass[twocolumn,10pt,superscriptaddress,aps,prd,floatfix]{revtex4-2}

\usepackage{textcomp}
\usepackage{color}
\usepackage{amsmath}
\usepackage{graphicx}
\usepackage{amssymb}
\usepackage{float}
\usepackage{mathbbold}

\usepackage[colorlinks=true,linkcolor=blue,citecolor=blue, linktocpage=true,breaklinks=true]{hyperref}

\newcommand \ket [1] {|{#1}\rangle}

\newcommand{\tr}{\mathrm{tr}}

\begin{document}

\title{Entanglement entropy production in deep inelastic scattering}

\author{Kun \surname{Zhang}}
\affiliation{Department of Chemistry, State University of New York at Stony Brook, Stony Brook, New York 11794, USA}
\author{Kun \surname{Hao}}
\email{haoke72@163.com}
\affiliation{Institute of Modern Physics, Northwest University, Xi'an 710069, China}
\affiliation{C.N. Yang Institute for Theoretical Physics, Stony Brook University, New York 11794, USA}
\author{Dmitri \surname{Kharzeev}}
\email{dmitri.kharzeev@stonybrook.edu}
\affiliation{Center for Nuclear Theory, Department of Physics and Astronomy, Stony Brook University, New York 11794, USA}
\affiliation{Physics Department, Brookhaven National Laboratory, Upton, NY 11973, USA}
\author{Vladimir \surname{Korepin}}
\email{vladimir.korepin@stonybrook.edu}
\affiliation{C.N. Yang Institute for Theoretical Physics, Stony Brook University, New York 11794, USA}
\affiliation{Department of Physics and Astronomy, Stony Brook University, New York 11794, USA}

\date{\today}

\begin{abstract}
	
	Deep inelastic scattering (DIS) samples a part of the wave function of a hadron in the vicinity of the light cone. Lipatov constructed a spin chain which describes the amplitude of DIS in leading logarithmic approximation. Kharzeev and  Levin proposed the entanglement entropy as an observable in DIS  [Phys. Rev. D 95, 114008 (2017)], and suggested a relation between the entanglement entropy and parton distributions.  Here we represent the DIS process as a local quench in the Lipatov's spin chain, and study the time evolution of the produced entanglement entropy. We show that the resulting entanglement entropy depends on time logarithmically, $\mathcal S(t)=1/3 \ln{(t/\tau)}$ with $\tau = 1/m$ for $1/m \le t\le (mx)^{-1}$, where $m$ is the proton mass and $x$ is the Bjorken $x$. The central charge $c$ of Lipatov's spin chain is determined here to be $c=1$; using the proposed relation between the entanglement entropy and parton distributions, this corresponds to the gluon structure function growing at small $x$ as $xG(x) \sim 1/x^{1/3}$.
	
\end{abstract}

\maketitle







\maketitle

\section{Introduction}

Fifty years ago, Balitsky, Fadin, Kuraev and Lipatov (BFKL)
set out a study of the high-energy behavior of the hadron scattering amplitude within perturbative QCD.  They identified the terms $(\alpha_s\ln s)^n$ (where $s$ is the squared centre-of-mass
energy and $\alpha_s$ is the strong coupling) resulting from the gluon ladders exchanged between the colliding hadrons. Since at high energies $\ln s$ is large, even at weak coupling it was necessary to resum the entire series of these 
leading logarithmic terms. The result was that
the total cross section grows as $s^{\alpha_{\rm BFKL} -1}$, where $\alpha_{\rm BFKL} > 1$ is the intercept of the resulting ``BFKL pomeron" \cite{BFKL,BFKL2,BFKL3,BFKL4}.

The growth of the cross section, and the corresponding increase of the gluon structure function at low Bjorken $x$, has been observed in deep inelastic scattering (DIS) at HERA \cite{H1,H1_2,H1_3,H1_4}, which excited interest in the studies of BFKL dynamics.
In a ground-breaking paper \cite{Lipatov94}, Lipatov discovered that in the leading logarithmic approximation (LLA), DIS can be effectively described by the XXX spin chain with zero spin.

\begin{figure}[H]
\centering
\includegraphics[width=5.0cm]{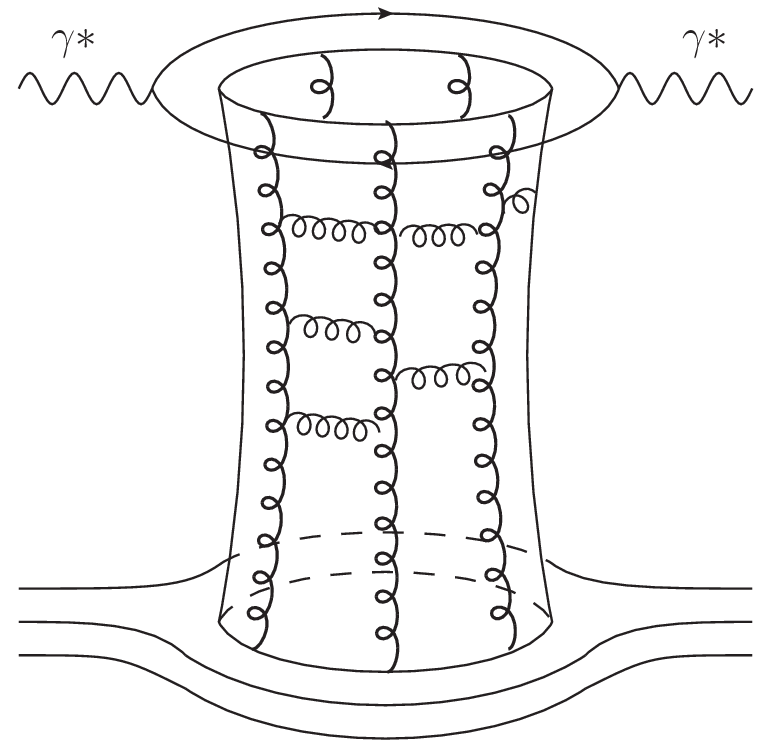}
\caption{Feynman diagram describing DIS at small Bjorken $x$. The virtual photon $\gamma*$ emitted by the scattered lepton (not shown) splits into a virtual quark-antiquark pair. The Reggeized  gluons are exchanged between the virtual quark-antiquark pair and the hadron.}
\label{fig1}
\end{figure}

At high energy, the scattering amplitudes in QCD are described by the exchange of gluons between the virtual quark-antiquark pair (resulting from the splitting of the virtual photon)  and the hadron. The gluons are  dressed by virtual gluon loops, which leads to their  "Reggeization". See Fig. \ref{fig1}. In the limit of large number of colors $N_c$ (with fixed $g^2 N_c$, where $g$ is the QCD coupling), the Hamiltonian describing the interactions of Reggeized gluons reduces to the sum of terms describing the  near-neighbor interactions, as  a Hamiltonian of a spin chain.
The chain was mapped to  the spin $(-1)$  \cite{Faddeev-Korchemsky1995}  
and to lattice nonlinear Schr\"odinger model \cite{Hao19}. 
Here  we will use the nonlinear Schr\"odinger (NLS) equation  \cite{Izergin-Korepin1982,Korepin-Izergin1982,Korepin-Izergin1982_2,Korepin93}  to describe the entanglement entropy evolution in DIS. In our treatment, we will rely on the conformal field theory (CFT) description of quantum lattice NLS.

Ideas of information theory find new applications in physics.  In particular, the quantum information approach to high-energy interactions was extended  in a recent paper \cite{Kharzeev21}, where it was argued that the phases of light cone wave functions cannot be measured in high-energy collisions -- therefore, the corresponding density matrix has to be averaged over the phase, with the corresponding Haar measure. This leads to the emergence of entanglement entropy describing the corresponding ``Haar scrambled" mixed states. The structure functions measured in DIS can be interpreted in terms of this entanglement entropy \cite{Kharzeev2017,Zhoudunming20}. To develop this description further and to describe the real-time evolution of the entanglement entropy in DIS, one needs to identify the physical excitations of the effective high-energy QCD Hamiltonian. This can be conveniently done in the spin chain case, with the help of the CFT description \cite{Cardy-Calabrese04,Cardy-Calabrese09,Korepin2004}. 
This motivates our study of entanglement entropy evolution in a local quench describing the DIS in the XXX spin chain with negative spin (which is equivalent to NLS).



The algebraic Bethe ansatz (quantum inverse scattering method) \cite{ABA,Korepin93} can be used for construction of the eigenfunctions of NLS.
Here we will apply this method to the XXX spin chain with spins $s=0$ and $-1$ describing high-energy QCD in the LLA.  Then we will explain the relation between the XXX spin chain with negative spin to the quantum NLS model.

The paper is organized as follows. In Sec. \ref{sec:Lipatov_spin_chain}, we present the construction of  eigenstates of Lipatov's spin chain by means of  the algebraic Bethe ansatz. In Sec. \ref{sec:QNS}, we will  map Lipatov's spin chain to the quantum lattice NLS model and study the thermodynamic limit of the system. In Sec. \ref{sec:entanglement_evolution}, we discuss the entanglement entropy evolution of Lipatov's spin chain after the local quench, and the corresponding evolution of the local operator entanglement. Sec. \ref{sec:conclusion} is the conclusion. Appendix \ref{appendix} provides an intuitive derivation of entanglement dynamics based on CFT. 


\section{Lipatov's spin chain}\label{sec:Lipatov_spin_chain}

The holomorphic multicolor QCD Hamiltonian \cite{Lipatov94,Faddeev-Korchemsky1995} describes the nearest neighbor interactions of $L$ particles (Reggeized gluons):
\begin{eqnarray}
H_L = \sum_{k=1}^L H_{k,k+1},
\label{H-gen}
\end{eqnarray}
with periodic boundary conditions $H_{L,L+1}=H_{L,1}$. To give a specific expression, let us introduce the holomorphic transverse coordinate $z_j$, and its corresponding momentum $P_j=i{\partial/\partial z_j}=i\partial_j$. Here $i=\sqrt{-1}$ is the imaginary unit. The local Hamiltonians are given by the equivalent representations
\begin{eqnarray}
H_{j,k}
&=&P_j^{-1}\ln(z_{jk})P_j
           +P_k^{-1}\ln(z_{jk})P_k
           +\ln(P_jP_k)+2\gamma_E
\nonumber
\\
&=&
2\ln(z_{jk})+(z_{jk})\ln(P_jP_k)(z_{jk})^{-1}+2\gamma_E,
\label{H2}
\end{eqnarray}
where $z_{jk}=z_j-z_k$, and $\gamma_E$ is the Euler constant. We have to put $j=k+1$ and substitute into (\ref{H-gen}).

Lipatov used a holomorphic representation of $SU(2)$
\begin{equation}
S_k^+ = z_k^2\partial_k-2s z_k ,
\qquad
S_k^- = -\partial_k ,
\qquad
S_k^z = z_k \partial_k - s,
\label{S}
\end{equation}
with $k=1,\ldots,L$. He then mapped DIS to a particular type of spin chain.
The definition of this chain is based on the existence of a fundamental matrix $R^{(s,s)}_{jk}(\lambda)$
which obeys the Yang-Baxter equation
\begin{eqnarray}
R^{(s,s)}_{jk}(\lambda)=f(s,\lambda)
\frac{\Gamma(i\lambda-2s)\Gamma(i\lambda+2s+1)}
                         {\Gamma(i\lambda-J_{jk})\Gamma(i\lambda+J_{jk}+1)}.
\end{eqnarray}
Here $f(s,\lambda)$ is a complex valued function (it normalizes the $R$ matrix), and  $\lambda$ is called the spectral parameter. The superscript $(s,s)$ means that both the auxiliary space and the quantum space have spin $s$. The operator $J_{jk}$ is defined in the space $V\otimes V$ as a solution of the operator equation,
\begin{eqnarray}
J_{jk}(J_{jk}+1) = 2\vec S_j\otimes\vec S_k+2s(s+1).
\label{J-operator}\end{eqnarray}
Everything commutes in this equation, so one can use Vieta's formula to solve this quadratic equation. The Hamiltonian of the XXX model with spin $s=0$ describes the interaction of nearest neighbors [see (\ref{H-gen})], which can be written as
\begin{align}
\label{H2xxx}
    H_{jk}=&\left.\frac{-1}{i}\frac{d}{d\lambda}
    \ln R^{(s=0)}_{jk}(\lambda)\right\vert_{{}_{\lambda=0}}, \\
    H_{jk}=&\psi(-J_{jk})+\psi(J_{jk}+1)-2\psi(1).
\end{align}
For simplicity, we apply the notation $H_{jk} = H_{j,k}$. Here $\psi(x)=d \ln\Gamma(x)/dx$, and $\psi(1)=-\gamma_E$ ($\gamma_E$ is the Euler constant). The operator $J_{jk}$ is a solution of (\ref{J-operator}) when $s=0$,
\begin{eqnarray}
J_{jk}(J_{jk}+1)=-(z_j-z_k)^2\partial_j\partial_k,
\label{spin-J}\end{eqnarray}
where we have to put $j=k+1$ to use in (\ref{H2xxx}).
This is a description of DIS in QCD by the $s=0$ spin chain.

After a similarity transformation, the spin $s=0$ model can be mapped to the $s=-1$ model. The latter can be easily solved by the algebraic Bethe ansatz method.
Thus the high-energy asymptotics in multicolor QCD is exactly solvable, and it has the same eigenvalues with that of the XXX spin $s=-1$ chain.

After finding the family of local integrals of motion and taking the XXX model of spin $s=-1$ into consideration, we can then apply the algebraic Bethe ansatz \cite{Tarasov-Takhtajan-Faddeev1983} in a standard procedure.

We define the auxiliary monodromy matrix
by taking the ordered product of the fundamental Lax operators $L^{(s,s)}_{f,k}(\lambda)=R^{(s,s)}_{f,k}(\lambda)$ \cite{Korepin93,Tarasov-Takhtajan-Faddeev1983} along the lattice (with both its auxiliary space and quantum space being spin $s$)
\begin{eqnarray}
T_f(\lambda) = L^{(s,s)}_{f,L}(\lambda)L^{(s,s)}_{f,L-1}(\lambda)\cdots L^{(s,s)}_{f,1}(\lambda).
\label{Tf}
\end{eqnarray}
The fundamental transfer matrix is the trace of the monodromy matrix over the auxiliary space,
\begin{eqnarray}
\tau(\lambda)=\mbox{tr}_f\, T_f(\lambda),\quad
[\tau(\lambda),\tau(\mu)]=0,
\label{tr}
\end{eqnarray}
i.e. these matrices commute with each other for different values of the spectral parameter.

On the other hand, if we choose the $L$ operator
\begin{eqnarray}
L^{({\frac 1 2},s)}_{a,k}(\lambda)=
\left(
  \begin{array}{cc}
    \lambda \mathbbold{1}_k+iS^z_k & iS^-_k \\
    iS^+_k & \lambda \mathbbold{1}_k-iS^z_k \\
  \end{array}
\right),
\end{eqnarray}
and define the transfer matrix
\begin{align}
\label{Ta}
    {t}(\lambda)=&\text{tr}_a[L^{({\frac 1 2},s)}_{a,L}(\lambda)\cdots L^{({\frac 1 2},s)}_{a,1}(\lambda)] \nonumber \\
    =&\text{tr}_a\left(\begin{array}{cc} A(\lambda) & B(\lambda) \\
    C(\lambda) & D(\lambda)
    \end{array}\right)\nonumber \\
    =&A(\lambda)+D(\lambda),
\end{align}
we get
\begin{eqnarray}
[t(\lambda),t(\mu)]=0,\quad[t(\lambda),\tau(\mu)]=0.
\end{eqnarray}
Both of the two transfer matrices $\tau(\lambda)$ and $t(\lambda)$ act on the full quantum space of the model and
commute with each other for different values of the spectral parameters.
One can get a family of mutually commuting conservation laws of the model.
The fundamental transfer matrix $\tau(\lambda)$ contains the local integrals of motion, including the Hamiltonian of the model.
In contrast, the operator $t(\lambda)$ allows one to construct their eigenstates by means of the Bethe ansatz.

The explicit forms of integrals of motions are given by \cite{Tarasov-Takhtajan-Faddeev1983}.
In particular, both the Hamiltonian of spin $s=-1$ and spin $s=0$ models can be obtained from the first order derivative of the transfer matrix $\tau$,
\begin{align}
\label{Hamiltonian}
&H_L^{(s=-1)}=\frac{-1}{i}\frac{d}{d\lambda}
\ln\tau^{(s=-1)}(\lambda)\bigg|_{\lambda=0}, \\
&H_L^{(s=0)}=\frac{-1}{i}\frac{d}{d\lambda}
\ln\tau^{(s=0)}(\lambda)\bigg|_{\lambda=0}.
\end{align}
Based on the relation between the Lax operators and the definition in Eq. (\ref{Ta}), the one-to-one correspondence between the XXX models of spin $s=-1$ and spin $s=0$ can be described by a similarity transformation (each local Hamiltonian of spin $s=0$ will be converted into a Hamiltonian with four nearest neighbor interactions):
\begin{equation}
H_L^{(s=-1)}=(z_{12}z_{23}\cdots{z}_{L1})^{-1}H_L^{(s=0)}z_{12}z_{23}\cdots{z}_{L1}.
\end{equation}
Thus the Hamiltonians of the two models have the same eigenvalues.

By using the explicit form of the spin operators (\ref{S}), one can find that for $s=-1$ the equations
\begin{equation}
S_k^+|\omega_k\rangle=0,\quad
S_k^z|\omega_k\rangle=-|\omega_k\rangle
\end{equation}
have the solution $|\omega_k\rangle=1/z^2_k$.
This allows us to construct the pseudovacuum state as
\begin{equation}
|\Omega\rangle=(z_1^2 z_2^2\cdots z_L^2)^{-1}.
\end{equation}
Then the Bethe states for spin $s=-1$ are given in terms of operator $B$ from (\ref{Ta}),
\begin{equation}
\ket{\hat\varphi_N(\{\lambda\})}=B(\lambda_1)B(\lambda_2)
\cdots B(\lambda_N)(z_1^2 z_2^2\cdots z_L^2)^{-1}.
\label{state}
\end{equation}
These are the {\it eigenvectors of Lipatov's spin chain}.
The eigenvalue of the transfer matrix $\tau(\lambda)$ as a function of spectral parameter $\lambda$ has the following form:
\begin{equation}
(\lambda-i)^L\frac{ Q(\lambda-i)}{Q(\lambda)}+(\lambda+i)^L\frac{ Q(\lambda+i)}{Q(\lambda)}\,,
\label{T-Q-xxx}
\end{equation}
with the function $Q(\lambda)$,
\begin{equation}
Q(\lambda)=\prod_{k=1}^N(\lambda-\lambda_k).
\end{equation}

The corresponding Bethe equation, determining the parameters $(\lambda_1,\ldots,\lambda_N)$, is
\begin{equation}
\left({\frac{\lambda_k+is}{\lambda_k-is}}\right)^L=\prod_{j=1,j\neq{k}}^N \frac{\lambda_k-\lambda_j+i}{\lambda_k-\lambda_j-i},
\end{equation}
with $k=1,\ldots,N$. Substitute $s=-1$, and then we have
\begin{equation}
\label{Beq}
\left({\frac{\lambda_k-i}{\lambda_k+i}}\right)^L=\prod_{j=1,j\neq{k}}^N \frac{\lambda_k-\lambda_j+i}{\lambda_k-\lambda_j-i}.
\end{equation}
These are periodic boundary conditions. In order to construct elementary excitations, we have to change to antiperiodic boundary conditions.
We remark that all the solutions $\lambda_k$ of the above Bethe equations are real numbers. This means that there is no bound state in this system.

The explicit expressions for the eigenvalues of integrals of motions
for arbitrary spin $s$ have been found in algebraic Bethe ansatz \cite{Tarasov-Takhtajan-Faddeev1983},
and we use these expressions for $s=-1$ to get the eigenvalues of the Hamiltonian
\begin{equation}
E\equiv\sum_{j=1}^N\frac{-1}{i}\frac{d}{d\lambda_j}
\ln\frac{\lambda_j+i}{\lambda_j-i}=\sum_{j=1}^N \frac{2}{\lambda^2_j+1},
\label{energy}\end{equation}
where $\{\lambda_j\}$ obey the Bethe equations (\ref{Beq})
for a fixed number of Reggeized gluons $L$. Thus, this relation yields the spectrum of the original holomorphic QCD model with Hamiltonian $H_L$.

\section{Quantum lattice nonlinear Schr\"odinger model}\label{sec:QNS}


Let us begin with a brief description of the quantum lattice nonlinear Schr\"odinger model. The quantum lattice NLS equation was introduced in \cite{Izergin-Korepin1982,Korepin-Izergin1982,Korepin93}. It is equivalent to the XXX spin chain with negative spin. Quantum lattice NLS is a chain of interacting harmonic oscillators. Let $\Psi_j^*$ and $\Psi_k$ be the canonical creation and annihilation operators of the harmonic oscillator:
\begin{equation}
[\Psi_j,\Psi_k^*]=\delta_{jk},
\end{equation}
and
\begin{equation}
\varrho_j=(1+{\frac{\kappa\Delta}{ 4}}\Psi_j^*\Psi_j)^{\frac 1 2}.
\end{equation}
Here $\delta_{jk}$ is the Kronecker delta function, $\kappa>0$ is the coupling constant \footnote{$\kappa$ is also a function of spin $s$: $\kappa=\big|{2/{s\Delta}}\big|$.} for NLS and $\Delta>0$ is a step of the lattice.
The operators
\begin{subequations}
\begin{align}
    {S_j}^x=&{\frac{i} {\sqrt{\kappa\Delta}}}(\Psi_j^*\varrho_j+\varrho_j\Psi_j),\\
    {S_j}^y=&{\frac{1}{\sqrt{\kappa\Delta}}}(\varrho_j\Psi_j-\Psi_j^*\varrho_j),\\
    {S_j}^z=&{\frac{-2}{\kappa\Delta}}(1+\frac{\kappa\Delta}{2}\Psi_j^*\Psi_j),
\end{align}
\end{subequations}
are the generators of an irreducible representation of $SU(2)$ algebra with a negative spin
\begin{equation}
 s=-\frac{2}{\kappa\Delta}.
 \label{spinValue}\end{equation}
In general, this $SU(2)$ representation \cite{Izergin-Korepin1982,Tarasov-Takhtajan-Faddeev1983} is infinite-dimensional, but for special (negative) values of $\Delta$ it can become finite-dimensional.

Let us focus on the correspondence between Bethe equations of the two models.
The Bethe roots $\lambda_{k}$ of the quantum lattice NLS model satisfy the following Bethe equations:
\begin{equation}
\label{Beq-NLS}
\left(\frac{1+i\lambda_{k}\Delta /2}{1-i\lambda_{k}\Delta /2} \right)^{L}=\prod^N_{j\neq k} \frac{\lambda_{k}-\lambda_{j}+i\kappa}{\lambda_{k}-\lambda_{j}-i\kappa}.
\end{equation}
Comparison of the above modified Bethe equations and Bethe equations (\ref{Beq}) shows the connections between the two models.
When we take coupling constant $\kappa=1$, and $\Delta=2$, the Bethe equations become
\begin{equation}
\label{Bethe-eq-NLS}
    (-1)^{L}\left(\frac{\lambda_{k}-i}{\lambda_{k}+i}\right)^{L}=\prod^{N}_{j\neq k}\frac{\lambda_{k}-\lambda_{j}+i}{\lambda_{k}-\lambda_{j}-i}.
\end{equation}
This means that the quantum lattice NLS model describes a more general XXX spin chain model with negative spin $s=-2/\kappa\Delta$, and holomorphic QCD is the special case with spin $s=-1$, $\Delta=2$ and coupling constant $\kappa=1$.


Based on Bethe equations (\ref{Beq}) and (\ref{Bethe-eq-NLS}) for the holomorphic QCD model (XXX with spin $s=-1$, also the quantum lattice NLS model), we have the logarithmic form Bethe equations.
We define each number $n$ (integer or half-integer) as a vacancy.
Among them, some vacancies corresponding to Bethe roots are called particles. Other free vacancies are called holes.
The number of vacancies is the sum of the number of particles and holes.

Differentiate the logarithmic Bethe equation with respect to $\lambda$, change the sum (in Bethe equations) to an integral, and one has the linear integral equation for the number (density) of vacancies $\rho_t(\lambda)$,
\begin{equation}
2\pi\rho_t(\lambda)=\int^{+\infty}_{-\infty}K(\lambda,\mu)\rho_p(\mu)d\mu+K(\lambda),
\label{rho-t-int}\end{equation}
with
\begin{equation}
K(\lambda,\mu)=\frac{2}{1+(\lambda-\mu)^2},\quad K(\lambda)=K(\lambda,0).
\end{equation}
Here $\rho_t(\lambda)$ is the sum of the numbers of particles $\rho_p(\lambda)$ and holes
$\rho_h(\lambda)$. Their proofs follow from \cite{Korepin93,Yang1968,Hao19}.

All $\lambda_j$ are different \cite{Korepin93} (Pauli principle in the momentum space). In the thermodynamic limit, the values of $\lambda_j$  condense and form a Fermi sphere.
Considering the grand canonical ensemble $E_h=E-h$ ($h$ is chemical potential) for small $h\rightarrow0^+$,
then all the vacancies inside the interval $(-\infty,-q]\cup[q,\infty)$ (called particles) are occupied by all the Bethe roots $\lambda_j$
(the density of holes $\rho_h(\lambda)=0$).
One can get a linear integral equation for $\rho_p(\lambda)$,
\begin{equation}
2\pi\rho_p(\lambda)=\left(\int^{-q}_{-\infty}+\int^\infty_{q}\right)K(\lambda,\mu)\rho_p(\mu)d\mu+K(\lambda).
\label{rho-p}\end{equation}

We define the dressed energy of elementary excitation $\varepsilon(\lambda)$ as the solution of the linear integral equation
\begin{align}
\varepsilon_0(\lambda)\equiv& \frac{2}{\lambda^2+1}-h\nonumber\\
=& \varepsilon(\lambda)-\frac{1}{2\pi}\left(\int^{-q}_{-\infty}+\int^{\infty}_{q}\right)
K(\lambda,\mu)\varepsilon(\mu)d\mu\;,
\end{align}
with condition
\begin{equation}
\varepsilon(q)=\varepsilon(-q)=0.
\end{equation}

\vskip0.3cm

In the correspondence between the structure functions and the entanglement entropy, the small $x$ behavior of the structure function is determined by the central charge of the effective CFT describing high-energy QCD \cite{Kharzeev2017}. It is thus important for us to evaluate the central charge of Lipatov's spin chain. 

The calculation of the central charge goes through the evaluation of finite size corrections. It can be calculated by means of the Bethe ansatz and CFT. The comparison gives the central charge.
The finite size correction to the ground state energy in continuous NLS was evaluated by means of the Bethe ansatz in Chapter 1 Sec. I.9 of the book \cite{Korepin93}.

For the current model, one can calculate the finite size correction to the ground state energy in the same way.
The ground state energy can be written as a summation with respect to the Bethe roots,
\begin{equation}
\frac{E}{L}={\frac{1}{L}}\sum_{j}\varepsilon_0\left({\lambda_j}\right).
\end{equation}
Using the Euler-Maclaurin formula for approximating sums by integrals,
one finally obtains
\begin{equation}
E=L\left(\int^{-q}_{-\infty}+\int^{\infty}_{q}\right)\varepsilon_0(\lambda)\rho(\lambda)d\lambda
-{\frac{\pi}{6L}}v_F+\text{h.o.c.}
\label{gse_fsc}\end{equation}
Here $v_F$ is the Fermi velocity, and h.o.c. means higher order corrections.

For unitary CFT, the central charge $c$, is the coefficient of the $1/ L$ term in the expansion of the ground state energy for $L\rightarrow\infty$,
\begin{equation}
E=L \tilde{\varepsilon} -c\frac{\pi v_F }{6L}+\text{h.o.c.}
\label{gse_cft}\end{equation}
See formula (0.1) in the Introduction of Chapter XVIII of the book \cite{Korepin93}.
For lattice NLS and more general cases, these can be obtained by the specification of Eq. (19) of the paper \cite{Izergin-Korepin-Reshetikhin-1989} (one has to put $M=1$).

Comparison of (\ref{gse_fsc}) with CFT (\ref{gse_cft}) shows that the {\it central charge of the corresponding Virasoro algebra is equal to one, $c=1$}.

\section{Time dependence of the entanglement entropy}\label{sec:entanglement_evolution}

\subsection{Entanglement entropy}

The entanglement entropy characterizes the lack of complete information about a subsystem (quantum fluctuation) when the total system is a known pure state. Suppose that the system is in the pure state, and consists of subsystems A and B; its state can be represented as
\begin{equation}
    |\Psi_{AB}\rangle = \sum_{j,k}\alpha_{j,k}|\psi_{A,j}\rangle\otimes |\psi_{B,k}\rangle,
\end{equation}
with the bipartition of Hilbert space $\mathcal H_{AB} = \mathcal H_A\otimes\mathcal H_B$ and $|\psi_{A(B)}\rangle\in\mathcal H_{A(B)}$. The pure state can be diagonalized in the subspace, given by
\begin{equation}
    |\Psi_{AB}\rangle = \sum_j \tilde\alpha_j |\psi_{A,j}\rangle\otimes |\psi_{B,j}\rangle,
\end{equation}
known as the Schmidt decomposition \cite{NC10}. The above decomposition naturally gives the density matrix of subsystem
\begin{equation}
\label{def:s_ab}
    \rho_{A(B)} = \tr_{B(A)}\rho_{AB} = \sum_j p_j |\psi_{A(B),j}\rangle\langle\psi_{A(B),j}|,
\end{equation}
with $p_j = |\tilde\alpha_j|^2$. Therefore the subsystems A and B are characterized by the same probabilistic distribution $p_j$, which is the signature of correlation. Then we consider the entanglement entropy, defined by
\begin{equation}
    \mathcal S_{A(B)} = -\sum_j p_j\ln p_j,
\end{equation}
which is the ignorance due to the correlation between subsystems A and B (we choose to use the natural log, corresponding to measuring information in ``nats"). In the basis-independent manner, the entanglement entropy has the form
\begin{equation}
    \mathcal S_A = -\tr \rho_A\ln\rho_A.
\end{equation}
Note that $\mathcal S_A = \mathcal S_B$, which demonstrates the correlation nature of the entanglement entropy. If the density matrix $\rho_A$ is given by the identity matrix, then the von Neumann entropy reaches the maximal value 
\begin{equation}
    \mathcal S_A = \ln d_A,
\end{equation}
with $d_A$ as the dimension of Hilbert space $\mathcal H_A$. Subsystems A and B are maximally entangled if $\mathcal S_A = \ln d_A$.

Quantum mutual information directly quantifies the amount of correlation between A and B, given by 
\begin{equation}
    \mathcal I(A;B) = \mathcal S_A+\mathcal S_B-\mathcal S_{AB}.
\end{equation}
In the case of a pure state, we have $\mathcal I(A;B) = 2\mathcal S_A$. Therefore the von Neumann entropy of a subsystem is equal to the half of mutual information between the two subsystems. 

\subsection{Evolution of entanglement entropy after quenches}

DIS probes a subregion A which in the rest frame of the proton is a tube with radius $1/Q$ and length $1/(mx)$ \cite{Gribov:1965hf}, where $Q$ is the  momentum transfer; $m$ is the proton mass; and $x$ is the Bjorken scaling variable. The  region inaccessible to the virtual photon is denoted as B. Since the proton represents a pure state which is an eigenstate of QCD Hamiltonian, the DIS probes a part of this state, and the unmeasured region has to be traced out. Then the entanglement entropy naturally arises in DIS \cite{Kharzeev2017}. If the entropy is indeed caused by the entanglement, then the entropy of rest of the nucleon should equal the entropy of parton distribution. The data from LHC (in the proton-proton collision) support the complementarity relation $\mathcal S_A = \mathcal S_B$ \cite{Zhoudunming20,H1:2020zpd,Kharzeev:2021yyf}. 
\vskip0.3cm

The (1+1)-dimensional systems described by CFT possess a universal scaling of entanglement entropy in the subsystem \cite{Holzhey94, Cardy-Calabrese04,Korepin2004}. Suppose that $\ell$ is the length of region A ($\ell\ll L$). Then its entanglement entropy is
\begin{equation}
\label{eq:S_A}
    \mathcal S_A = \frac c 3\ln \frac{\ell}{\epsilon},
\end{equation}
with the central charge $c$ and the ultraviolet cutoff $\epsilon$ (the resolution scale). For spin chains, this logarithmic formula was rigorously proven in \cite{Jin04} (Fredholm determinants and the Riemann-Hilbert problem were used). Therefore, Lipatov's spin chain with central charge 1 predicts the logarithmic state entanglement. In an effective (1+1)-dimensional model of QCD evolution \cite{Mueller94,Mueller95}, the entanglement entropy is found to be $\mathcal S_A = \delta \ln(1/x)$ where the constant $\delta$ thus describes the growth of structure function $xG(x) \sim (1/x)^\delta$ at small $x$.

In the target rest frame, the cutoff $\epsilon=1/m$ is given by the proton's Compton wavelength and $\ell = 1/(mx)$ is the longitudinal distance probed in DIS. The correspondence between the central charge of the CFT and the intercept of the gluon structure function is thus $\delta = c/3$ \cite{Kharzeev2017}. The experimental data indicate $\delta\approx 0.3$ (see e.g.  \cite{Albacete11,Hentschinski13,Iancu15}), which supports the CFT description of Lipatov's spin chain with the central charge $c=1$ determined above. 

 The DIS process can be understood as a local quench on the ground state of the proton. The quench causes a local excitation, which propagates in time, and at time $\simeq 1/(mx)$ saturates. It is likely that entanglement causes an effective thermalization of the system (thermalization through entanglement) \cite{Kaufman16}. Entanglement thermalization in the proton-proton scattering has been discussed in \cite{Baker18}. Here we argue that the evolution of entanglement in DIS can be described by the local quench of Lipatov's spin chain or the corresponding CFT with central charge $c=1$.
\vskip0.3cm

Calabrese and Cardy have studied the entanglement evolution after a local quench, based on the CFT method \cite{Calabrese07}. The initial state of the evolution corresponds to the ground states of regions A and B separately. Therefore the translation invariance is broken. There is a defect on the boundary of A and B (local excitation), and the entanglement entropy increases logarithmically: 
\begin{equation}
\label{eq:S_A_t/3}
    \mathcal S_A(t) = \frac c 3\ln\frac t \tau,
\end{equation}
with the characteristic time $\tau$. This formula can also be derived by comparison of the theory of classical shock waves with the CFT; see Appendix \ref{appendix}.
The quasiparticle excitations are emitted only from the defect point, and therefore the entanglement entropy undergoes a logarithmic increase. A linear increase would require a global quench (sudden change of the entire Hamiltonian) \cite{Calabrese05}. 

The characteristic time $\tau$ is determined by the boundary condition between A and B. It is independent of the central charge and is beyond the CFT description.  In the case of DIS, in the target rest frame it is given by the proton's Compton wavelength $\tau = 1/m$. 
The time evolution of entanglement entropy is thus given by
\begin{equation}
    \mathcal S_A(t) = \frac c 3 \ln mt,
\end{equation}
which agrees with the CFT description of Lipatov's spin chain. After the critical time $t_c = 1/(mx)$, the entanglement saturates, and the region A probed by DIS becomes maximally entangled with the remaining part of the proton (which is not probed by the virtual photon). 



\subsection{Evolution of operator entanglement entropy}

Entanglement evolution after a local quench characterizes the spreading of entanglement from a local region. In a complementary Heisenberg picture, the operators evolve as $O(t) = e^{iHt}Oe^{-iHt}$. Consider the operator space $\mathcal H'_L = \text{End}(\mathcal H_L)$ with the Hilbert-Schmidt inner product $\langle O_j|O_k\rangle = \tr(O_j^\dag O_k)$, $O_{j(k)}\in \text{End}(\mathcal H_L)$. Similar to the state entanglement, the space can be divided into two regions A and B. Then the operator has the Schmidt decomposition
\begin{equation}
    \frac{O}{\sqrt{\langle O|O\rangle}} = \sum_j \sqrt{\chi_j} O_{A,j}\otimes O_{B,j},
\end{equation}
with the eigenvalues $\chi_j$ and the orthonormal bases $\langle O_{A(B),j}|O_{A(B),k}\rangle = \delta_{jk}$. Similar to the state entanglement entropy defined in Eq. (\ref{def:s_ab}), the operator space entanglement entropy (OSEE) is defined as
\begin{equation}
    \mathcal S(O) = -\sum_j \chi_j\ln\chi_j.
\end{equation}
The evolution of OSEE in terms of local operators also characterizes the entanglement spreading in the system. OSEE was first introduced in \cite{Zanardi01}. Then it was reintroduced in \cite{Prosen07,Prosen07_2} to study the simulation of quantum dynamics. It is suggested that the OSEE grows at most logarithmically in integrable systems, while chaotic systems have linear increases. 

Numerical studies (based on the density matrix renormalization group) have shown that the XXX spin chains with positive spin  ($s = 1/2$ and $s=1$) have a logarithmic increase of OSEE (for local operators) \cite{Muth11,Alba19,Alba20}. However, the prefactor in front of the logarithm is different from the state entanglement evolution. In terms of the local projection $O = 1/2-S^z_k$, numerical results show that
\begin{equation}
    \mathcal S(O(t)) \propto \frac 2 3\ln t,
\end{equation}
in the XXX-1/2 spin chain \cite{Alba20}. Note that different local operators may have different prefactors. We argue that the Lipatov's spin chain (XXX spin chain with $s=-1$) has a similar logarithmic increase of OSEE. It is consistent with the state entanglement evolution after local quench, given by Eq. (\ref{eq:S_A_t/3}). Besides, a logarithmic increase is a general feature of integrable systems. However, the evolution of OSEE for local operators cannot be described by CFT in general \cite{Dubail17}.

Another interesting observation is the OSEE evolution in the quantum cellular automaton. Cellular automaton has both discrete space and time.  Quantum cellular automaton has the unitary evolution and  is a natural language for quantum computation. Different rules of cellular automaton have different names. OESS for the local operator also grows logarithmically (the same as the XXZ-$1/2$ spin chain) in the quantum cellular automaton rule 54 \cite{Alba19,Alba20}. Such a sublinear increase of entanglement entropy suggests its efficient simulation on quantum computers \cite{Eisert21}. A logarithmic increase of OSEE in Lipatov's spin chain suggests that DIS can be efficiently simulated on quantum computers.  We leave these simulations for the future.




\section{Conclusions}\label{sec:conclusion}

In high-energy QCD, the scattering amplitude of DIS in the LLA has been described by Lipatov in terms of an integrable spin chain model. We mapped this model to the quantum lattice NLS model. We then derived  the eigenfunctions  by means  of the algebraic Bethe ansatz.  After evaluation of finite size corrections, we have concluded that the Virasoro algebra (describing an effective CFT) has the central charge equal to one, $c=1$. Based on the CFT description, we found that the time evolution of entanglement entropy after local quench is logarithmic
\begin{equation}
\label{eq:result}
    \mathcal S_A(t) ={\frac{1}{3}} \ln\frac t \tau,  
\end{equation}
with $\tau = 1/m$ for $1/m \le t\le (mx)^{-1}$, where $m$ is the proton mass and $x$ is the Bjorken $x$.  The integrable system also has the logarithmic evolution of OSEE. This suggests that the DIS process can be efficiently simulated on quantum computers. 


\section*{Acknowledgments}

The authors thank Xuanhua Wang and Professor Vincenzo Alba for the helpful discussions. This material is based upon work supported by the U.S. Department of Energy, Office of Science, National Quantum Information Science Research Centers, Co-design Center for Quantum Advantage (C2QA) under Contract No. DE-SC0012704. K.H. was supported by the National Natural Science Foundation of China (Grants No. 11805152, No. 12047502, and No. 11947301), Shaanxi Natural Science Fundamental Research Program
No. 2021JCW-19, and Shaanxi Key Laboratory for
Theoretical Physics Frontiers in China. The work of D.K. was supported by the U.S. Department of Energy, Office of Science Grants No. DEFG88ER40388 and No. DE-SC0012704.

\appendix

\section{CFT description of entanglement entropy dynamics}\label{appendix}
\renewcommand{\theequation}{A.\arabic{equation}}



Let us consider the entropy of a block of spins (an interval of an infinite system). At positive temperature, thermal fluctuations dominate. A theory of classical shock waves shows that   after local quench  the  entropy is a linear function of time:
\begin{equation}
\mathcal S(t)=\frac{2\pi c}{3} T t, 
\end{equation}
where $T$ is the temperature.
The coefficient can be found in \cite{Cardy-Calabrese09}; 
see for example their formula (80).  The shock wave changes the density of the entropy.



Now let us consider entanglement entropy  at zero temperature, which is quantum.  The entropy is some function $f$ of time,
\begin{equation}
\mathcal S(t)=f(t).
\end{equation}
Conformal mapping shows that for positive $T>0$
\begin{equation}
\mathcal S_T(t)=f\left(\frac{v}{\pi{T}}\sinh\frac{\pi{T}}{v}(x+vt)\right),
\end{equation}
where $v$ is velocity. We can put $x=0$ and consider the limit of large time.
The result for the entanglement entropy at time $t$ is
\begin{equation}
\mathcal S_T(t)=f\left(\exp[\pi T (t-t_0)]\right), 
\end{equation}
where $t_0$ is an inessential constant. Now we have two expressions for the entropy for positive temperature
\begin{equation}
\mathcal S_T(t)=\frac{2\pi c}{3} T t =f\left(\exp[\pi T (t-t_0)]\right).
\end{equation}
This means that we have found  the function, and it is given by
\begin{equation}
f(t)=\frac{2c}{3} \ln t + \text{const}.
\end{equation}
Each end of the block contributes equally, so for the local quench we get 
\begin{equation}
\mathcal S(t)=\frac{c}{3} \ln t + \text{const}.
\end{equation}
It agrees with the entanglement entropy evolution after local quench (with one point of defect) calculated in \cite{Calabrese07}.





\providecommand{\noopsort}[1]{}\providecommand{\singleletter}[1]{#1}%

\end{document}